\documentclass[aps,amsmath,amssymb,twocolumn,showpacs]{revtex4-1}

\usepackage{graphicx}
\usepackage{dcolumn}
\usepackage{bm}
\usepackage{color}

\begin{document}

\title{Length dependence of the thermal conductance of alkane-based
  single-molecule junctions: An ab-initio study}

\author{J. C. Kl\"ockner$^{1}$}
\email{Corresponding author: Jan.Kloeckner@uni-konstanz.de}
\author{M. B\"urkle$^{2}$}
\author{J. C. Cuevas$^{3}$}
\author{F. Pauly$^{1}$}

\affiliation{$^{1}$Department of Physics, University of Konstanz, D-78457 Konstanz, Germany}

\affiliation{$^{2}$National Institute of Advanced Industrial Science and Technology
(AIST), Umezono 1-1-1, Tsukuba Central 2, Tsukuba, Ibaraki 305-8568,
Japan}

\affiliation{$^{3}$Departamento de F\'{\i}sica Te\'orica de la Materia Condensada and Condensed Matter
Physics Center (IFIMAC), Universidad Aut\'onoma de Madrid, E-28049 Madrid, Spain}

\date{\today}

\begin{abstract}
Motivated by recent experiments, we present here a systematic ab-initio study
of the length dependence of the thermal conductance of single-molecule
junctions. We make use of a combination of density functional theory with
non-equilibrium Green's function techniques to investigate the length
dependence of the phonon transport in single alkane chains, contacted with
gold electrodes via both thiol and amine anchoring groups. Additionally, we
study the effect of the substitution of the hydrogen atoms in the alkane
chains by heavier fluorine atoms to form polytetrafluoroethylenes. Our results
demonstrate that (i) the room-temperature thermal conductance is fairly
length-independent for chains with more than 5 methylene units and (ii) the
efficiency of the thermal transport is strongly influenced by the strength of
the phononic metal-molecule coupling. Our study sheds new light onto the
phonon transport in molecular junctions, and it provides clear guidelines for
the design of molecular junctions for thermal management.
\end{abstract}

\pacs{63.22.-m, 65.80.-g, 68.65.-k, 73.63.Rt}


\maketitle

\section{Introduction}

Heat conduction by phonons is a fundamental physical process with great
relevance in numerous problems in science and engineering as well as in
technological applications \cite{Minnich2014}. In the context of macroscale
systems, phonon conduction is, to some extent, well understood thanks to the
use of semiclassical theoretical methods such as the Boltzmann transport
equation \cite{Ziman-book} or molecular-dynamics-based approaches
\cite{Minnich2014}, which are often used in combination with ab-initio
techniques for the determination of the phonon spectra and coupling
constants. However, with the advent of novel nanoscale systems and devices it
has become possible to study phonon conduction in a new regime, where the
system size can be smaller than the inelastic mean free path for phonons (due
to phonon-phonon interactions), and in some cases even smaller than the
corresponding elastic mean free path leading to quasi-ballistic transport
\cite{Cahill2002,Pop2010,Luo2013,Cahill2014}. In this new regime, phonon
conduction has to be described by purely quantum mechanical means, which poses
a very interesting and challenging problem for the theory.

In this work we are interested in the heat conduction via phonons in
single-molecule junctions, which constitute the ultimate limit of
miniaturization of electronic devices. These nanojunctions have emerged in
recent years as ideal platforms to study the quantum theories that describe
both the charge and energy transport in nanoscale devices
\cite{Cuevas2010}. In particular, recent experimental advances have made it
possible to investigate different aspects of energy and heat conduction in
molecular junctions such as thermoelectricity
\cite{Reddy2007,Kim2014,Rincon-Garcia2016}, Joule heating \cite{Lee2013}, and
thermal conductance \cite{Segal2016}. These advances stimulate the development
of fully ab-initio methods, which in turn should elucidate the main aspects
and physical mechanisms of phonon conduction in atomic-scale junctions. In
this sense, the goal of this paper is to use an ab-initio approach, based on a
combination of density functional theory (DFT) and non-equilibrium Green's
function (NEGF) techniques, to study one of the most fundamental questions in
the context of phonon transport in atomic-scale systems, namely the length
dependence of the thermal conductance of a molecular chain.

The theoretical discussion of the thermal conductance of a linear one-dimensional 
(1D) chain has a long history \cite{Lepri2003,Dhar2008,Li2012}. The general conclusion 
is that in long ideal chains (of more than 100 identical units) exhibiting nonlinear 
interactions, momentum conservation and no disorder, the thermal conductance decays
algebraically as $\kappa_{\rm ph} \propto L^{\alpha-1}$, where $L$ is the
chain length and the exponent $\alpha$ is found to be $\alpha = 1/3$. Although
these results are of fundamental interest, it is not clear that they are very
relevant for actual experiments on nanoscale systems due to the long chain lengths 
required.

In the context of molecular junctions, the issue of the length dependence of
the thermal conductance has already been addressed both experimentally and
theoretically. Often the molecules of choice have been alkane chains
\cite{Wang2007,Majumdar2015}, which are saturated molecules, whose electrical
properties have been widely studied in the context of molecular electronics
\cite{Cui2001,Wang2003,Haiss2004,Li2006,Gonzalez2006,
  Fujihira2006,Jang2006,Chen2006,Venkataraman2006,Park2007,Li2008,Gonzalez2008,Martin2008,Akkerman2008}.
From the experimental side, the length dependence of the thermal conductance
of alkane-based molecular junctions was investigated by Wang \emph{et
  al.}\ \cite{Wang2006} making use of alkanedithiol self-assembled monolayers
(SAMs) that were sandwiched between Au and GaAs electrodes. It was found that
the thermal conductance of junctions with 8, 9, and 10 CH$_2$ units did not
depend significantly on the molecule length. More recently, Meier \emph{et
  al.}\ \cite{Meier2014} reported thermal conductance measurements of
monothiolated alkane monolayers, self-assembled on Au(111) surfaces as a
function of their length (ranging from 2 to 18 methylene units). Making use of
a scanning thermal microscope with a Si tip, it was found that the thermal
conductance first increases for short chains, reaching a maximum for 4 units,
and then it exhibits an (arguable) slow decay for lengths above 8 units.  From
the theory side, Segal \emph{et al.}\ \cite{Segal2003} made use of
semi-empirical methods to predict that the thermal conductance of alkane
chains approaches a constant value for more than 20 CH$_2$ units for a weakly
coupled junction, while it decreases inversely proportional to the length for
the strongly coupled case. On the other hand, Duda \emph{et
  al.}\ \cite{Duda2011}, using a diffusive transport model combined with
Hartree-Fock calculations of the vibrational modes of alkane chains, suggested
that the thermal conductance should be fairly length-independent for chains
with more than 5 CH$_2$ units. In the only ab-initio study of this issue that
we are aware of, Sadeghi \emph{et al.}\ \cite{Sadeghi2015} explored the phonon
transport in alkane-based single-molecule junctions with 2, 4, 8, and 16
CH$_2$ units and found a very pronounced decay of the thermal conductance for
the longest molecular length. These seemingly contradictory theoretical
results call for new inspections of the fundamental issue of the length
dependence of the thermal conductance of molecular junctions.

In this work we present a systematic ab-initio study with no adjustable
parameters of the phonon transport in alkane-based single-molecule junctions,
but discuss also electronic contributions to the heat transport. Making use of
a combination of DFT and NEGF techniques, we analyze the thermal conductance
of Au-alkane-Au molecular junctions as a function of the length of the alkane
chains ranging from 2 to 30 methylene units. Moreover, in order to elucidate
the role of the metal-molecule interface, we investigate two different
anchoring groups, namely thiol (SH) and amine (NH$_2$). Additionally, we
determine the impact of the substitution of the H atoms in the alkane chains
by F atoms for both types of anchoring groups. The ensemble of our results
shows that the room-temperature phonon thermal conductance of all investigated
junctions is fairly length-independent for chains with more than 5 CH$_2$
units. This statement also holds for the total thermal conductance that
consists of phononic and electronic contributions, since we show that the
electronic contribution plays a significant role only for the shortest chain
lengths below 4 units. Moreover, we demonstrate that the relative efficiency
of the phonon heat conduction is mainly, although not exclusively, dictated by
the strength of the metal-molecule phononic coupling. Our results yield novel
insights into the heat conduction in molecular wires and provide clear
predictions that will surely be possible to test in the near future due to the
recent rapid advances in the experimental techniques.

The rest of the manuscript is organized as follows. First, in
section~\ref{sec-Methods} we briefly describe the theoretical approach
employed to obtain the results presented in this work. Then, in
section~\ref{sec-Results} we describe and analyze the main results of this
manuscript on phonon heat transport. Section~\ref{sec-Further} is devoted to
additional discussions of our results such as electronic contributions to the
thermal conductance and to the comparison with published experimental and
theoretical work. Finally, we present in section~\ref{sec-Conclusions} our
conclusions.

\section{Theoretical approach} \label{sec-Methods}

Our primary goal is to compute the thermal conductance due to phonon transport
in single-molecule junctions. For this purpose, we make use of the
first-principles formalism developed by us and reported in
Ref.~[\onlinecite{Burkle2015}]. Our approach is based on a combination of DFT
and NEGF that allows us to compute both the electronic and phononic
contributions to all basic linear response transport properties of a nanoscale
system. In what follows, we briefly describe the main features of our method
and refer the reader to Ref.~[\onlinecite{Burkle2015}] for further details.

\subsection{Contact geometries, electronic structure, and vibrational properties}

The first step in our modeling is the construction of the molecular junction
geometries. We use DFT to compute equilibrium geometries through total energy
minimization and to describe their electronic structure. Vibrational
properties of the optimized equilibrium contacts are subsequently obtained in
the framework of density functional perturbation theory. We use both
procedures as implemented in the quantum chemistry software package TURBOMOLE
6.5 \cite{TURBOMOLE,Deglmann2002,Deglmann2004}, employ the PBE
exchange-correlation functional \cite{Perdew1992,Perdew1996}, the basis set
def2-SV(P) \cite{Weigend2005} and the corresponding Coulomb fitting basis
\cite{Weigend2006}. To make sure that the vibrational properties,
i.e.\ vibrational energies and force constants, are accurately determined, we
use very stringent convergence criteria. Thus, total energies are converged to
a precision of better than $10^{-9}$~a.u., while geometry optimizations are
performed until the change of the maximum norm of the Cartesian gradient is
below $10^{-5}$~a.u.

\subsection{Phonon transport}

We compute the phononic contribution to the heat conductance within the
framework of the Landauer-B\"uttiker picture, i.e., we ignore inelastic and
anharmonic effects that are expected to play a minor role in short molecular
junctions. Within this picture, the heat current due to phonons can be
expressed as \cite{Rego1998,Mingo2003,Yamamoto2006}
\begin{equation}
  J_{\rm ph}(T_{\rm L},T_{\rm R}) = \dfrac{1}{h} \int_{0}^{\infty} dE \, E \tau_{\rm ph}(E)
  \left[ n(E,T_{\rm L}) - n(E,T_{\rm R})\right], \label{eq-Qph}
\end{equation}
where $\tau_{\rm ph}(E)$ is the phonon transmission, $n(E,T)=[\exp(E/k_{\rm
    B}T)-1]^{-1}$ the Bose function describing the phonon occupation in the
left (L) or right (R) electrode, and $T_X$ is the temperature in electrode
$X=\text{L},\text{R}$. In this work, we focus on the linear response regime in
which $J_{\rm ph} = -\kappa_{\rm ph} \Delta T$ is proportional to the
temperature difference $\Delta T=T_{L}-T_{R}$ and the phonon thermal
conductance is given by
\begin{equation}
  \kappa_{\rm ph}(T) = \frac{1}{h} \int_{0}^{\infty} dE \,E \tau_{\rm ph}(E)
  \frac{\partial n(E,T)}{\partial T}. \label{eq-kph}
\end{equation}

The phonon transmission appearing in the previous equations can be determined
with the help of Green's function techniques
\cite{Mingo2003,Yamamoto2006,Burkle2015}. Briefly, the starting point is the
description of the phonons or vibrational modes of the junction within the
harmonic approximation. In this approximation, the phonon Hamiltonian for
small displacements $\{Q_{\xi}\}$ of the atoms around their equilibrium
positions $\{R_{\xi}^{(0)}\}$ adopts the following form 
\begin{equation}
  \hat H = \frac{1}{2} \sum_{\xi} \hat{p}_{\xi}^{2} + \frac{1}{2\hbar^{2}} 
  \sum_{\xi\chi} \hat{q}_{\xi} \hat{q}_{\chi} K_{\xi\chi},
\end{equation}
where we have introduced mass-weighted displacement operators $\hat{q}_{\xi} =
\sqrt{M_{\xi}}\hat{Q}_{\xi}$ and mass-scaled momentum operators
$\hat{p}_{\xi}=\hat{P}_{\xi}/\sqrt{M_{\xi}}$ as conjugate variables. These
variables obey the following commutation relations:
$[\hat{q}_{\xi},\hat{p}_{\chi}]=\mbox{i}\hbar\delta_{\xi\chi}$ and
$[\hat{q}_{\xi},\hat{q}_{\chi}]=[\hat{p}_{\xi},\hat{p}_{\chi}]=0$, where
$\xi=(i,c)$ denotes a Cartesian component $c=x,y,z$ of atom $i$ at position
$\vec{R}_{i}=\vec{R}_{i}^{(0)}+\vec{Q}_{i}$.  The phonon system is
characterized by its dynamical matrix $K_{\xi\chi} = \hbar^{2}
\mathcal{H}_{\xi\chi} /\sqrt{M_{\xi}M_{\chi}}$, which is the mass-weighted
Hessian of the DFT total ground state energy with respect to the Cartesian
atomic coordinates, $\mathcal{H}_{\xi\chi} = \partial_{\xi\chi}^{2}E$. These
harmonic force constants are computed within density functional perturbation
theory.

The use of a local displacement basis enables the partitioning of the
dynamical matrix into three parts, a central (C) scattering region, and the
two semi-infinite L and R electrodes, i.e.\
\begin{equation}
  \boldsymbol{K}=\left(\begin{array}{ccc}
    \boldsymbol{K}_{\mathrm{LL}} & \boldsymbol{K}_{\mathrm{LC}} & \mathrm{\boldsymbol{0}}\\
    \boldsymbol{K}_{\mathrm{CL}} & \boldsymbol{K}_{\mathrm{CC}} & \boldsymbol{K}_{\mathrm{CR}}\\
    \boldsymbol{0} & \boldsymbol{K}_{\mathrm{RC}} & \boldsymbol{K}_{\mathrm{RR}}
  \end{array}\right). \label{eq-Kmatrix}
\end{equation}
Notice that there is no direct coupling between L and R. The phonon
transmission can be expressed as \cite{Mingo2003,Asai2008}
\begin{equation}
  \tau_{\rm ph}(E) = \mathrm{Tr} \left[ \boldsymbol{D}_{\mathrm{CC}}^{\textrm{r}}(E) 
    \boldsymbol{\Lambda}_{\textrm{L}}(E) \boldsymbol{D}_{\mathrm{CC}}^{\textrm{a}}(E)
    \boldsymbol{\Lambda}_{\textrm{R}}(E)\right], \label{eq-tauph}
\end{equation}
where $\boldsymbol{D}_{\mathrm{CC}}^{\textrm{r,a}}(E)$ are the retarded and
advanced phonon Green's functions of the central region that can be computed
by solving the following Dyson equation
\begin{equation}
  \boldsymbol{D}_{\mathrm{CC}}^{\mathrm{r}}(E) = \left[\left(E+i\eta
    \right)^{2}\boldsymbol{1} - \boldsymbol{K}_{\textrm{CC}} -
    \boldsymbol{\Pi}_{\textrm{L}}^{\mathrm{r}}(E) -
    \boldsymbol{\Pi}_{\textrm{R}}^{\mathrm{r}}(E) \right]^{-1}
\end{equation}
with
$\boldsymbol{D}_{\mathrm{CC}}^{\mathrm{a}}(E)=\boldsymbol{D}_{\mathrm{CC}}^{\mathrm{r}}(E)^{\dagger}$
and an infinitesimal quantity $\eta>0$. On the other hand, the
linewidth-broadening matrices
\begin{equation}
  \boldsymbol{\Lambda}_{X}(E)= i
  \left[\boldsymbol{\Pi}_{X}^{\mathrm{r}}(E)-\boldsymbol{\Pi}_{X}^{\mathrm{a}}(E)\right]
\end{equation}
are related to the corresponding contact self-energies
\begin{equation}
  \boldsymbol{\Pi}_{X}^{\mathrm{r}}(E) = \boldsymbol{K}_{\textrm{C}X} \boldsymbol{d}_{XX}^{\mathrm{r}}(E)
  \boldsymbol{K}_{X\mathrm{C}},
\end{equation}
describing the coupling between the central region C and electrode $X$. In the
expressions $\boldsymbol{d}_{XX}^{\mathrm{r}}(E)$ is the surface Green's
function of lead $X=\text{L},\text{R}$ and
$\boldsymbol{\Pi}_{X}^{\textrm{a}}(E)=\boldsymbol{\Pi}_{X}^{\textrm{r}}(E)^{\dagger}$.

In analogy with the concept of conduction channels in coherent electron
transport \cite{Cuevas2010}, it is interesting to decompose the total phonon
transmission in terms of individual transmission coefficients of the different
phononic scattering eigenfunctions that contribute to the phonon conduction at
each energy. The total phonon transmission in Eq.~(\ref{eq-tauph}) can be
rewritten as a sum of independent contributions
\begin{equation}
  \tau_{\rm ph}(E) = \mbox{Tr} \{ \boldsymbol{t}_{\rm ph} (E)
  \boldsymbol{t}_{\rm ph}^{\dagger} (E) \} = \sum_n \tau_{{\rm ph},n}(E) ,
\end{equation}
where $\boldsymbol{t}_{\rm ph}(E) = \boldsymbol{\Lambda}^{1/2}_{\textrm{R}}(E)
\boldsymbol{D}_{\mathrm{CC}}^{\textrm{r}}(E) \boldsymbol{\Lambda}^{1/2}_{\textrm{L}}(E)$
is the phonon transmission matrix and $\tau_{{\rm ph},n}(E)$ are the energy-dependent 
eigenvalues of $\boldsymbol{t}_{\rm ph} \boldsymbol{t}_{\rm ph}^{\dagger}$.

To conclude this discussion, let us remark that to compute the different parts
of the dynamical matrix in Eq.~(\ref{eq-Kmatrix}) we follow the strategy of
our cluster-based approach for charge quantum transport \cite{Pauly2008}. In
this approach, we first compute the dynamical matrix for an extended central
cluster including the molecule and large parts of the leads. Subsequently, we
extract from there the matrices $\boldsymbol{K}_{\mathrm{CC}}$,
$\boldsymbol{K}_{\mathrm{XC}}$.  On the other hand, the surface Green's
functions of the electrodes $\boldsymbol{d}_{XX}^{\mathrm{r}}(E)$ are obtained
by extracting bulk force constants from a separate calculation of a big
cluster of several hundred atoms and then using these extracted parameters in
combination with a decimation technique to describe the surface of a
semi-infinite perfect crystal (see Ref.~[\onlinecite{Burkle2015}] for
details). In this way, we achieve a consistent, parameter-free ab-initio
description of the phonon system in nanoscale devices.

\subsection{Electronic transport}

For completeness, in this work we also investigate the electronic contribution
to the thermal conductance. In this way we can assess the relative importance
of electrons and phonons in the studied molecular junctions. Similar to the
phononic case, we assume that the electronic transport is dominated by elastic
tunneling processes. Under these circumstances the electronic contribution to
the thermal conductance in the linear response regime can be computed within
the Landauer-B\"uttiker formalism. It is given by \cite{Sivan1986,Cuevas2010}
\begin{equation}
  \kappa_{\rm el}(T) = \frac{2}{hT} \left( K_{2}(T)- \frac{K_{1}(T)^{2}}{K_{0}(T)}
  \right), \label{eq-kel}
\end{equation}
where the $K_n(T)$ coefficients are defined as
\begin{equation}
  K_n(T) = \int^{\infty}_{-\infty} dE \, \tau_{\rm el}(E) \left(-\tfrac{\partial
    f(E)}{\partial E}\right)(E-\mu)^{n},
  \label{eq-Kn}
\end{equation}
$\tau_{\rm el}(E)$ is the energy-dependent electron transmission and
$f(E,\mu,T) = \left\{ \exp[(E-\mu)/k_{\mathrm{B}}T]+1\right\} ^{-1}$ is the
Fermi function. Here, the chemical potential $\mu\approx E_{\textnormal{F}}$
is approximately given by the Fermi energy $E_{\textnormal{F}}$ of the Au
electrodes.

We have computed the electron transmission making use of our DFT-NEGF
formalism implemented in TURBOMOLE and explained in detail in
Ref.~\onlinecite{Pauly2008}. It is worth stressing that in order to correct
for the known inaccuracies in DFT related to quasiparticle energies, we have
made use of the DFT+$\Sigma$ approach \cite{Quek2007}, which was implemented
in our method as explained in Ref.~[\onlinecite{Zotti2014}].

\section{Results} \label{sec-Results}

To investigate the length dependence of the thermal conductance of
single-molecule junctions, we study alkanes chains of different length with an
even number $n$ of methylene (CH$_2$) units ranging between 2 and 30.  These
saturated molecules are electrically rather insulating, which ensures that the
thermal transport is dominated by phonons (see below). To elucidate the role
of the anchoring or terminal group in the length dependence of the phonon
transport, we analyze two standard groups in molecular electronics, namely
thiol (S) and amine (NH$_2$) groups. Moreover, we study the impact of the
substitution of the H atoms in the alkane chains by heavier F atoms to
investigate polytetrafluoroethylenes (PTFEs) with both thiol and amine
groups. Thus, in an attempt to draw general conclusions, we investigate four
different families of molecules attached in all cases to gold electrodes. 

\begin{figure}[t]
\begin{center} \includegraphics[width=0.9\columnwidth,clip]{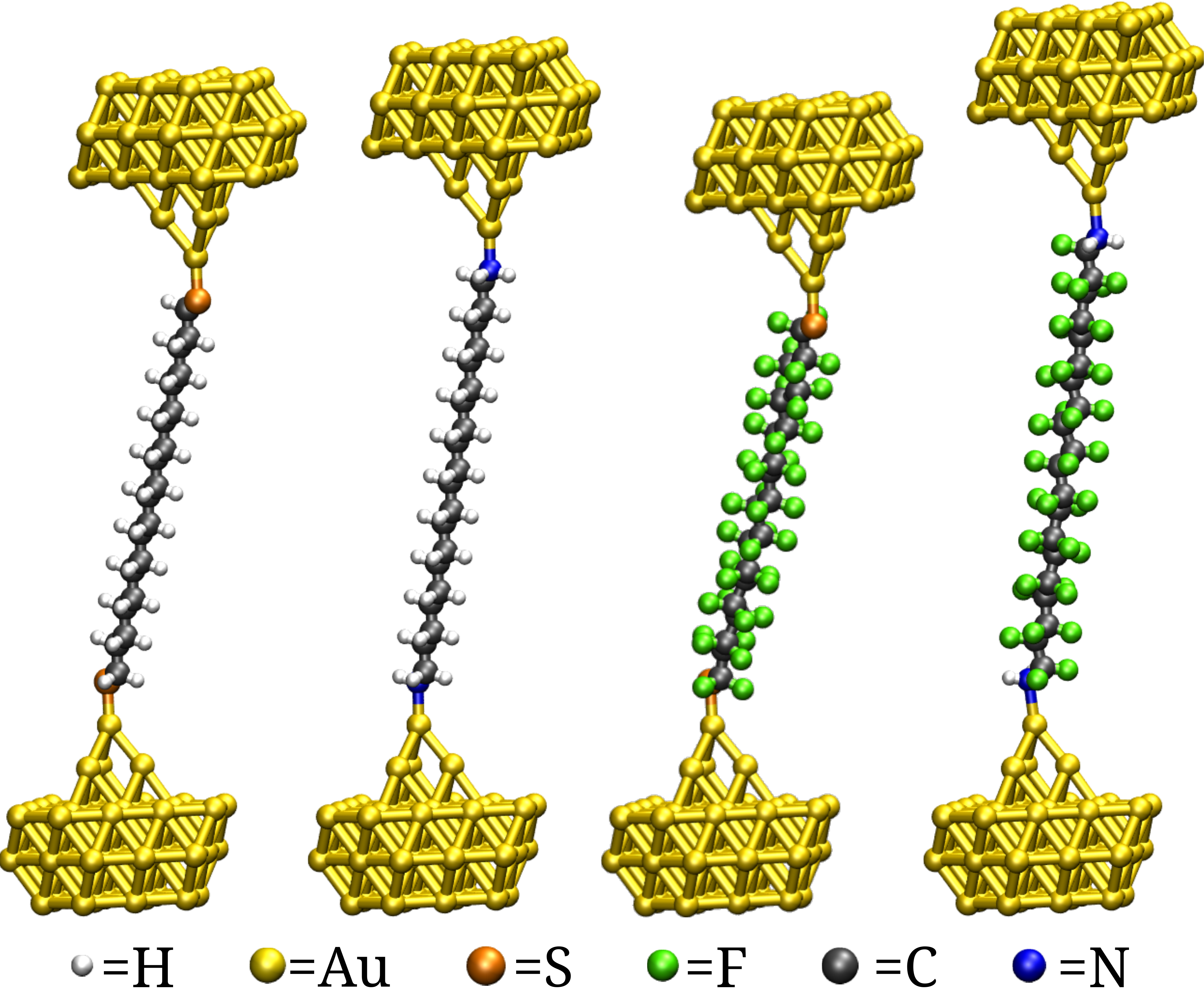} \end{center}
\caption{(Color online) Examples of the four types of molecular junctions
  studied in this work.  The four molecules have the structure
  $X$-(C$Y_2$)$_n$-$X$, where the terminal or anchoring group $X$ is either S
  or NH$_2$, while the $Y$ atom in the repeated unit is either H or F.  The
  integer number $n$ denotes the number of C$Y_2$ segments in the
  molecules. In all cases, the electrodes are made of gold and the contact
  geometries are such that molecules are bonded to the electrodes in an atop
  position via the anchoring groups.}
\label{fig-geometries}
\end{figure}

In what follows, we shall focus on the binding geometries illustrated in
Fig.~\ref{fig-geometries}, where the molecules are bonded to the electrodes in
an atop position via the corresponding anchoring group. Let us stress that in
all cases the molecular junctions were carefully optimized to find the minimum
energy configuration. Moreover, in order to establish a meaningful comparison
between the different compounds, special care was taken to avoid both strain
effects and the appearance of defects in the molecular chains.  In this sense,
the alkane chains remain linear in our case, while the introduction of F atoms
causes the carbon atoms to deviate from this linear structure, leading to more
disordered PTFE chains. Independent of this, the distance between the carbon
atoms stays nearly constant in all chains with a value of around 1.5 \AA.

\begin{figure}[t]
\begin{center} \includegraphics[width=\columnwidth,clip]{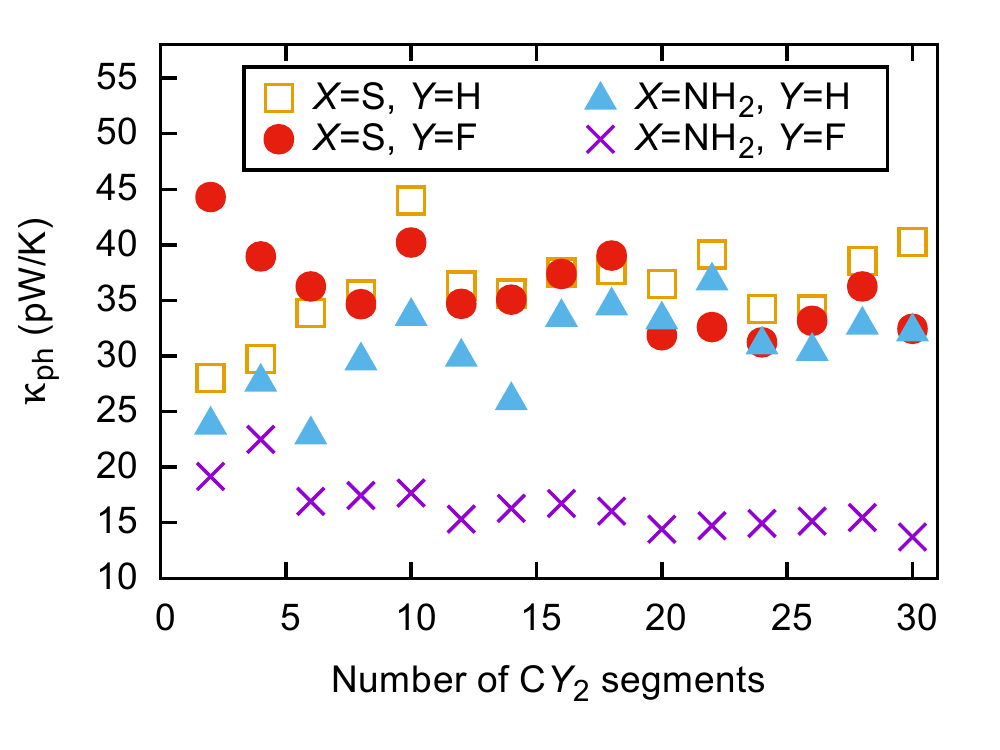} \end{center}
\caption{(Color online) Room-temperature ($T=300$~K) phonon thermal
  conductance as a function of the number of C$Y_2$ units
  ($Y=\text{H},\text{F}$) in the molecule for both anchoring groups, thiol and
  amine. The distance between neighboring C atoms in the chains of around
  1.5~\AA \ allows to translate the number of units to a corresponding
  distance.}
\label{fig-kph}
\end{figure}

We summarize in Fig.~\ref{fig-kph} the main result of this work, namely the
room-temperature phononic thermal conductance for the four types of molecular
junctions as a function of the number of C$Y_2$ segments
($Y=\text{H},\text{F}$) ranging from 2 to 30, which corresponds to a maximum
length of around 4.5 nm. The first thing to mention is that the conductance
values range from 15 to 45 pW/K. Second, for chains with more than 5 segments
the conductance exhibits small variations, but it is basically
length-independent, irrespective of the molecular family. Third, the molecules
with amine anchoring groups typically exhibit a lower conductance than the
corresponding dithiolated ones. This is particularly evident in the case of
the PTFE-diamine molecules, which exhibit thermal conductances that are about
a factor of 2 smaller than those of the other families. Overall, these results
show that the phononic thermal conductance of alkane-based single-molecule
junctions is rather insensitive to the molecular length for up to 30 segments,
which is a signature of ballistic phonon transport.

\begin{figure}[t]
  \begin{center} \includegraphics[width=\columnwidth,clip]{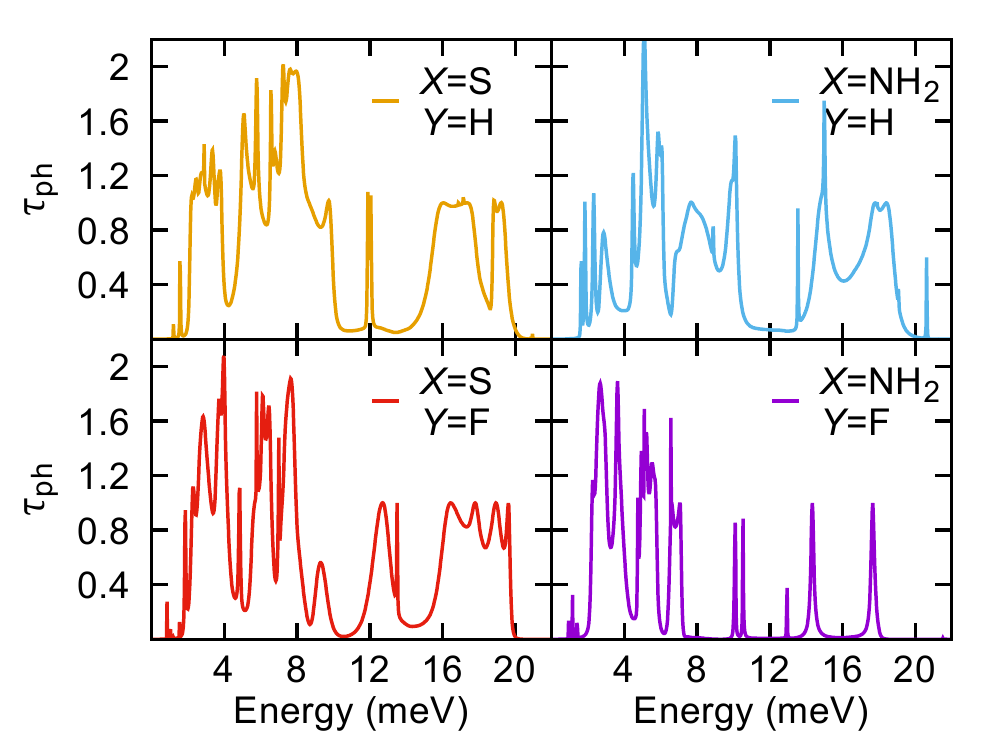} \end{center}
  \caption{(Color online) Phonon transmission as a function of energy for
    junctions containing molecules with 10 C$Y_2$ units.}
  \label{fig-Tph}
\end{figure}

To understand these results, let us first analyze the energy dependence of the
phonon transmission.  An example for the four types of molecules with $n=10$
C$Y_2$ segments is shown in Fig.~\ref{fig-Tph}. First of all, notice that the
transmission is different from zero only in an energy region between 0 and 20
meV, which is determined by the phonon density of states of the gold
electrodes (see Ref.~[\onlinecite{Burkle2015}]). Second, the most obvious
feature in these results is the fact that for the molecules with the amine
group the transmission spectra exhibit narrower peaks, which explains the
lower conductance obtained for these molecules. (This is particularly well
pronounced for the PTFE-diamine ones.) This strongly suggests that the
phononic metal-molecule coupling for the amine group is weaker than for
thiol. On the other hand, although the fluorinated molecules are expected to
exhibit more vibrational modes in the transport window than the alkane chains
due to the larger mass of the F atoms, one does not observe significant
differences.

\begin{figure}[b]
\begin{center} \includegraphics[width=\columnwidth,clip]{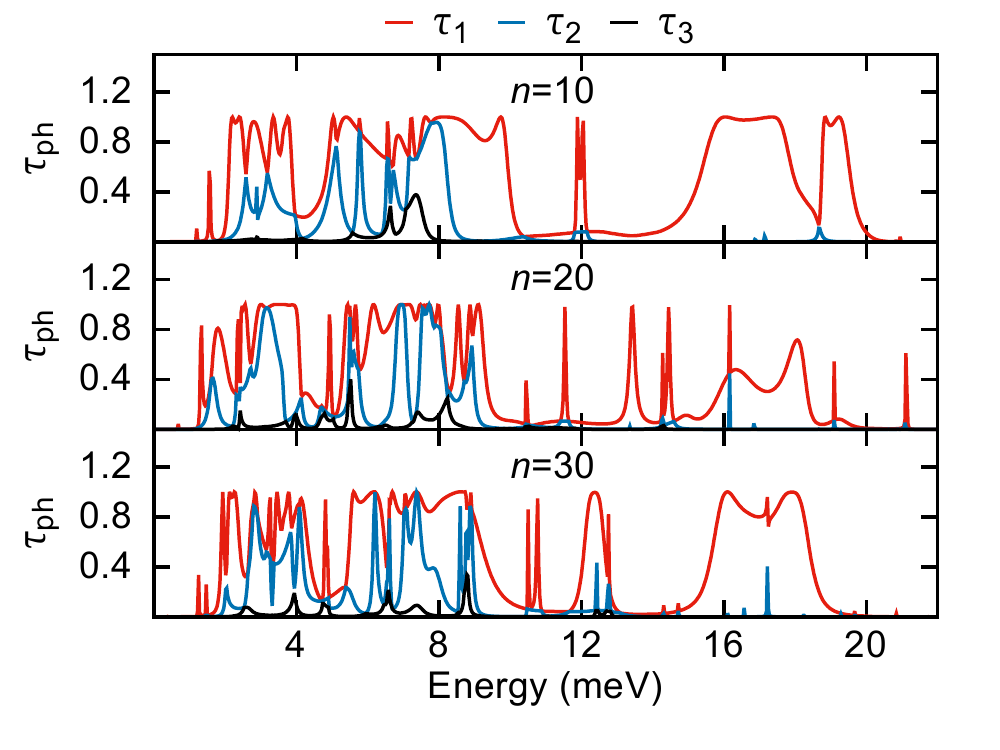} \end{center}
\caption{(Color online) Individual phonon transmission coefficients as a
  function of the energy for three alkanedithiols with different lengths (10,
  20, and 30 methylene units).  We show the coefficients of the three most
  transmissive channels.}
\label{fig-taun}
\end{figure}

To further clarify how the elastic phonon transport takes place, we show in
Fig.~\ref{fig-taun} the largest individual phonon transmission coefficients
for molecular junctions with alkanedithiols of three different lengths
($n=10$, $20$, and $30$ methylene units). The first thing to notice is that
the total transmission is dominated by a single or two phonon channels, with a
third one giving only a small contribution. The number of channels is
controlled in this case by the rather linear molecules, and the three channels
correspond to the three polarizations of the vibrational modes. It is
important to realize that this number of channels cannot be altered by the
number of vibrational modes in the molecular chains, which only determine the
actual values of the transmission coefficients. More important for our
discussion of the length dependence is the fact that the dominant channel is
fully open (transmission equal to 1) over a wide range of energies,
irrespective of the length of the molecule. This is indeed the true signature
of ballistic phonon transport, which is realized here with alkane and
alkane-related molecular chains.


\begin{figure}[tb]
\begin{center} \includegraphics[width=0.9\columnwidth,clip]{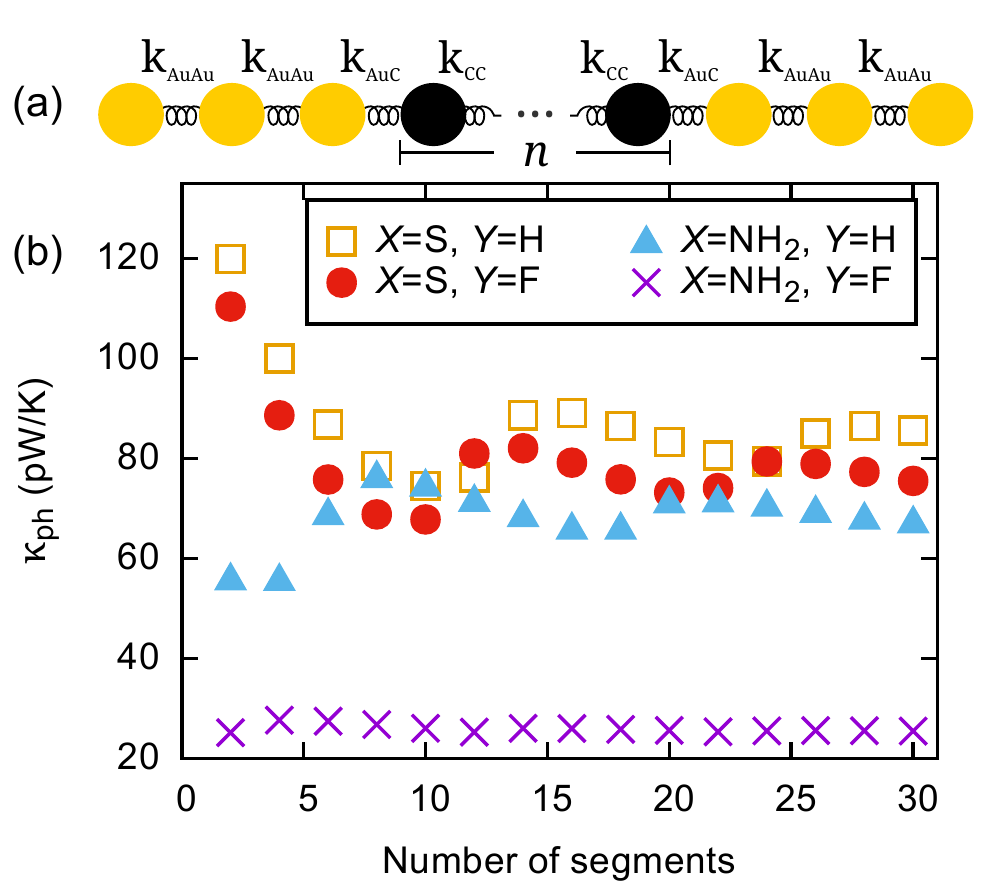} \end{center}
\caption{(Color online) (a) Schematic representation of the 1D model used to
  understand our findings.  The meaning of the different parameters is
  explained in the text. (b) Results obtained with the 1D model for the
  room-temperature phonon thermal conductance as a function of the number of
  C$Y_2$ units ($Y=\text{H},\text{F}$) in the molecule for both anchoring
  groups, thiol and amine.}
\label{fig-1Dmodel}
\end{figure}

To gain further insight into our ab-initio results, we have developed a simple
1D model that is schematically represented in Fig.~\ref{fig-1Dmodel}(a). In
the following we will use the notation $k_{ij}$ for elements of the dynamical
matrix coupling atoms $i$ and $j$ in the toy model as compared to $K_{ij}$ in
the full ab-initio results.  In the model we consider only nearest-neighbor
couplings between segments in the molecular chain, $k_{\rm CC}$, and the leads
are modeled as 1D Au chains with an analogous nearest-neighbor coupling,
$k_{\rm Au Au}$. Finally, the metal-molecule coupling is described by a single
constant, $k_{\rm Au C}=k_{{\rm Au} X}\sqrt{M_X/M_{\rm C}}$ with
$X=\text{S},\text{N}$. We extracted these parameters from the DFT calculations
as the highest eigenvalue of $K_{ij}$, see Eq.~(\ref{eq-Kmatrix}).  These
parameters depend on the molecular species, but not on the molecular length,
and we summarize their values in Table \ref{table}.

\begin{table}[b]
\begin{tabular}{lcccc}
\hline 
$X$  & $Y$  & $k_{\mathrm{Au}X}$  & $k_{\mathrm{CC}}$ &$k_{\rm Au Au}$ \tabularnewline
\hline 
S  & H  & 485 & 6150 & 100 \tabularnewline
S & F  & 426 & 4850 & 100 \tabularnewline
N  & H  & 218 & 6150 & 100 \tabularnewline
N & F  & 55  & 4850 & 100 \tabularnewline
\hline 
\end{tabular}\caption{Parameters for the 1D model schematically represented in
  Fig.~\ref{fig-1Dmodel}(a). All elements $k_{ij}$ of the dynamical matrix are
  given in units of meV$^2$. Notice that the same value of $k_{\rm Au Au}$ was
  used in all cases. \label{table}}
\end{table}

Using these parameters, we computed the corresponding thermal conductance with
the Green's function method described in section \ref{sec-Methods} and the
results are displayed in Fig.~\ref{fig-1Dmodel}(b). As one can see, this
simple model is able to reproduce all the salient features of our ab-initio
results in Fig.~\ref{fig-kph}. In particular, it nicely reproduces the fact
that the amine-terminated molecules exhibit a lower thermal conductance, which
is especially evident in the case of the PTFE-diamine chains. Now we can
confirm that this lower thermal conductance is due to a weaker coupling to the
leads (see values of $k_{\mathrm{Au}X}$ in Table~\ref{table}). The strongly reduced coupling for PTFE-diamine chains as compared to their alkane-diamine counterparts
results in practice from the larger distance (about 0.15 \AA) between the Au electrodes
and the N atoms in these molecular junctions. Let us also mention
that we attribute the higher conductance values obtained with the 1D model as compared to the ab-initio results to
the fact that we extract the parameters $k_{ij}$ from the largest eigenvalues of the
dynamical submatrices $K_{ij}$.

\section{Further discussion} \label{sec-Further}

\begin{figure}[tb]
\begin{center} \includegraphics[width=\columnwidth,clip]{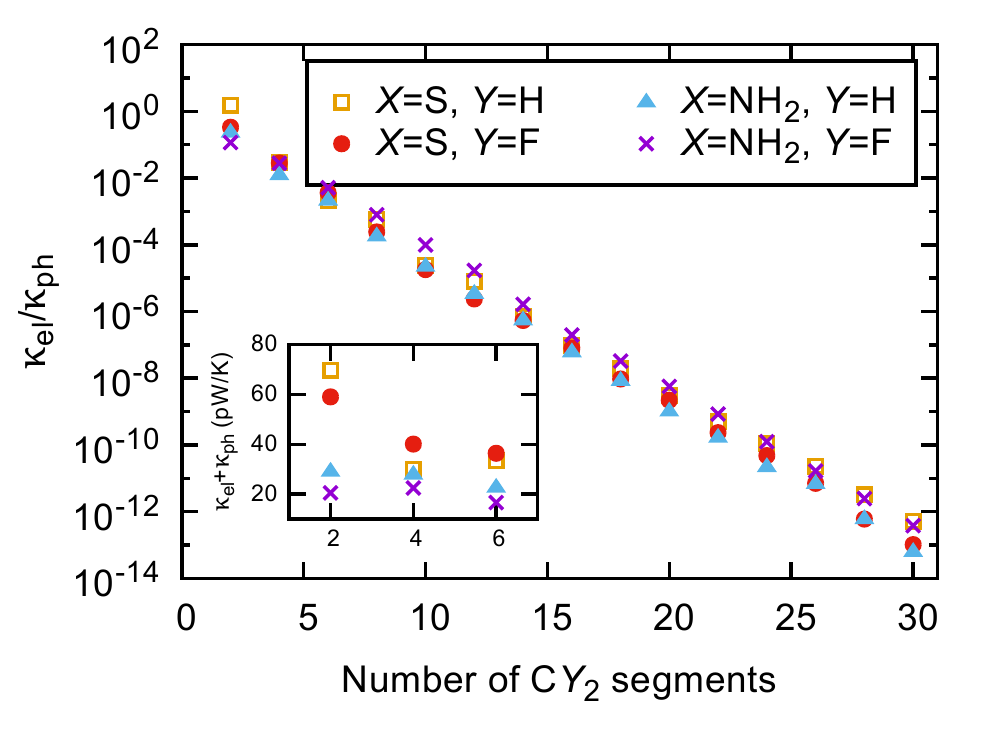} \end{center}
\caption{(Color online) Ratio of the electronic ($\kappa_{\rm el}$) and
  phononic ($\kappa_{\rm ph}$) room-temperature thermal conductance as a
  function of the molecular length for the 4 different types of molecules
  studied in this work. In the inset the thermal conductance due to both
  electronic and phononic contributions is displayed for short chain lengths.}
\label{fig-ratio}
\end{figure}

So far we have focused on the phononic contribution to the thermal conductance
and one may wonder whether the electrons play any role. Alkanes are known to
be poor electrical conductors
\cite{Cui2001,Wang2003,Haiss2004,Li2006,Gonzalez2006,
  Fujihira2006,Jang2006,Chen2006,Venkataraman2006,Park2007,Li2008,Gonzalez2008,Martin2008,Akkerman2008}.
Thus, on the basis of the Wiedemann-Franz law \cite{Cuevas2010,Burkle2015},
one therefore does not expect the electrons to give a significant contribution
to the thermal conductance. To check this, we have studied the electronic
contribution to the thermal conductance, $\kappa_{\rm el}$, in all of our
molecular junctions using the ab-initio methodology briefly described in
subsection \ref{sec-Methods}.C. Using $T=300$~K, we show in
Fig.~\ref{fig-ratio} the results for the ratio between the electronic and
phononic thermal conductances as a function of the molecular length for the
four families of molecules investigated here. As one can see, when molecular
chains have more than 4 segments, the electronic contribution is
negligible. Notice that the exponential decay of the conductance ratio is a
simple consequence of the exponential decay of the electrical conductance with
length in the off-resonant transport situation. It follows from the
proportionality of $\kappa_{\rm el}$ and the electrical conductance and a
rather length-independent $\kappa_{\rm ph}$.

\begin{figure}[bt]
\begin{center} \includegraphics[width=0.9\columnwidth,clip]{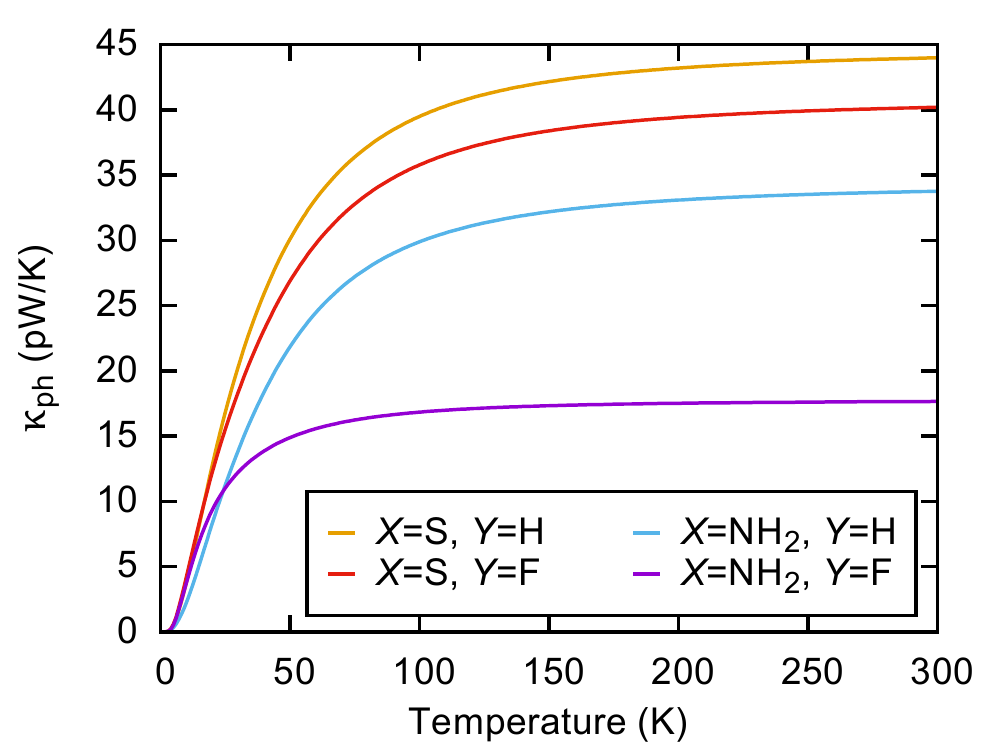} \end{center}
\caption{(Color online) Phonon thermal conductance as a function of the
  temperature for junctions containing molecules with 10 C$Y_2$ units.}
\label{fig-Tdep}
\end{figure}

In the previous section we discussed the results for the thermal conductance
at room temperature.  For completeness, we now briefly address the issue of
the temperature dependence of the phonon transport. Fig.~\ref{fig-Tdep}
displays the temperature dependence from 0 to 300 K of the phonon thermal
conductance for junctions with the four molecular species and featuring $n=10$
C$Y_2$ segments. This temperature dependence is relatively insensitive to the
molecular length and thus, the results of Fig.~\ref{fig-Tdep} are
representative of the four molecular species studied in this work. As one can
see, the thermal conductance raises abruptly at low temperatures, it tends to
saturate, depending on the molecule, above approximately 100 to 200 K, and
around room temperature it is fairly constant. This overall behavior simply
reflects the fact that for temperatures above the Debye temperature of gold,
all the phonons of the metal electrodes as well as the vibrational modes of
the molecule with energies within the transport window are thermally occupied,
while below this temperature the higher-lying modes are only partially
occupied and the thermal conductance hence becomes sensitive to the
temperature.

Let us now discuss the comparison with existent results. From the theory side
our results, showing a thermal conductance relatively insensitive of the
molecular length, are qualitatively compatible with the findings of Segal
\emph{et al.}\ \cite{Segal2003} for weakly coupled alkane chains. This is also
the case for the results of Duda \emph{et al.}\ \cite{Duda2011}. Our results
are also in agreement with the molecular dynamics calculations performed by
Luo and Lloyd \cite{Luo2010}, which studied octanedithiol SAMs sandwiched
between gold electrodes and estimated a thermal conductance per molecule of 43
pW/K. However, our results are clearly at variance with the strong decay for
the longest molecule reported by Sadeghi \emph{et al.} \cite{Sadeghi2015},
which indeed is the only ab-initio study published to date on the length
dependence of the thermal conductance of alkane single-molecule junctions. The
reason for the discrepancy is unclear to us, but it is worth pointing out that
those authors only analyzed a very limited number of molecules (with 2, 4, 8,
and 16 methylene units) and they used a different anchoring group
(hydrobenzothiophene).

With respect to existent experiments on the length dependence, we cannot
establish a direct comparison, but our length independence is compatible with
the experiments of Wang \emph{et al.}\ \cite{Wang2006} in junctions based on
alkanedithiol SAMs sandwiched between Au and GaAs electrodes. On the other
hand, the experiments of Meier \emph{et al.}\ \cite{Meier2014}, which
represent the most systematic study of the length dependence of the thermal
conductance of alkane junctions to date, exhibit some basic differences with
the junctions studied in this work. First, the tip electrode was made of Si.
Second, the alkanes were monothiolated. Third, the temperature difference used
in the experiment was between 200 and 300 $^{\rm o}$C, which is most likely
beyond the linear response regime addressed in our work. Finally, in this
experiment the number of contacted molecules is not directly determined, and
it is only inferred with the help of tip models. Having said all that, let us
mention that the estimated conductance values per molecule lie in the range of
our calculated values and a careful inspection of the experimental data shows
that the conductance is indeed fairly independent of the molecular length for
chains with more than 8 methylene units. Anyway, a rigorous comparison with
experiment to settle the issue of the length dependence of the thermal
conductance requires true single-molecule experimental techniques, and our
results provide clear predictions that we hope will be tested experimentally
in the near future.

Related to the role of the anchoring group, let us mention that Losego
\emph{et al.}\ \cite{Losego2012} investigated experimentally with the
time-domain thermoreflectance technique its impact on heat transport in SAMs
of alkanes contacted to quartz and gold films. They observed that the
replacement of amines as the binding group to the Au film by thiols led to a
60\% increase of the thermal conductance per unit area for chains with 11
methylene units. This is consistent with the general trend found here in the
sense that the amine-terminated molecules exhibit in general a lower thermal
conductance than their thiolated counterparts.

Let us conclude this section by stressing that in spite of the fact that we
are using an approach, where anharmonic effects are not taken into account,
the finding of a length-independent thermal conductance, as reported, here is
by no means trivial. The intrinsic disorder in the molecular chains, which is
present e.g.\ in the PTFE chains, can in principle lead to a diffusive
transport regime, where the conductance decays linearly with length, or
ultimately to an Anderson-localized regime, where the conductance is
exponentially suppressed with length. In our study we see that for chains of
up to 30 segments (with a length of up to 4.5 nm) the conductance exhibits a
weak length dependence, indicating that the elastic mean free path is larger
than the system size. It would be interesting to study the limits of this
quasi-ballistic transport, and we hope to address this issue in the near
future. Let us also emphasize that Segal \emph{et al.}\ \cite{Segal2003}
showed that anharmonicity in the alkane chains is rather weak and it is not
expected to play a crucial role in the range of lengths studied in this
work. Therefore, our approach is well justified.


\section{Conclusions} \label{sec-Conclusions}

In summary, we have presented a systematic ab-initio study of the length
dependence of the thermal transport through alkane-based single-molecule
junctions. In particular, we have investigated the role of the anchoring group
in this length dependence and we have compared the results for alkane chains
with those for PTFE chains, which are obtained from the alkanes by
substituting the H atoms by F atoms. By investigating the phonon transport for
chains with up to 30 segments, we find that the phonon thermal conductance is
rather insensitive to the molecular length, irrespective of the
molecule. These results strongly suggest that the phonon transport in these
molecular wires is quasi-ballistic. On the other hand, our analysis of the
role of the anchoring group shows that these groups ultimately determine the
efficiency of the phonon transport and, in particular, we find that the thiol
group leads to higher conductance values than the amine group. Electronic
contributions to the thermal conductance do not modify these conclusions,
since we demonstrate that they are negligible for alkanes with 4 or more
segments. Overall, our results provide clear predictions that we hope will be
tested experimentally in the near future given the rapid advances in the field
of nanothermometry. Moreover, our work sheds new light on the thermal
transport mechanisms in molecular wires and provides clear guidelines for the
design of molecular junctions for thermal management.

\section{Acknowledgments}

J.C.K.\ and F.P.\ gratefully acknowledge funding from the Carl Zeiss
foundation and the Junior Professorship Program of the Ministry of Science,
Research, and the Arts of the state of Baden-W\"urttemberg.  M.B.\ was
supported by a Grant-in-Aid for Young Scientists (Start-up) (KAKENHI
\#15H06889) from the Japan Society for the Promotion of Science and
J.C.C.\ through the Spanish Ministry of Economy and Competitiveness (Contract
No.\ FIS2014-53488-P) and thanks the German Research Foundation (DFG) and
Collaborative Research Center (SFB) 767 for sponsoring his stay at the
University of Konstanz as Mercator Fellow. An important part of the numerical
modeling was carried out on the computational resources of the bwHPC program,
namely the bwUniCluster and the JUSTUS HPC facility.


\end{document}